\documentclass[twocolumn,eqsecnum,superscriptaddress,showpacs]{revtex4}
\usepackage{bm}

\begin{document}
\title{Mistery of Real Scalar Klein-Gordon Field }
\author{Takehisa \textsc{Fujita}} \email{fffujita@phys.cst.nihon-u.ac.jp}
\author{Seiji \textsc{Kanemaki}} \email{kanemaki@phys.cst.nihon-u.ac.jp}
\author{Atsushi \textsc{Kusaka}} \email{kusaka@phys.cst.nihon-u.ac.jp}
\author{Sachiko \textsc{Oshima}} 
\affiliation{Department of Physics, Faculty of Science and Technology, 
Nihon University, Tokyo, Japan }%

\date{\today}

\begin{abstract}
The mystery that the real scalar Klein-Gordon field has vanishing 
current densities is resolved. The scalar field is shown to be 
a complex field due to the condition of possessing a proper non-relativistic 
limit. Like the Schr\"odinger field, one component complex Klein-Gordon field  corresponds 
to a boson with one flavor, and therefore there exists no physical real scalar field. 
As a good example, we present the Schwinger model Hamiltonian which is naturally 
described by the complex scalar field with one flavor. The non-existence of the real 
scalar field indicates that the Higgs mechanism should be reconsidered.

\end{abstract}


\maketitle

\section{ Introduction}

The Klein-Gordon equation has a history of 80 years, and has been repeatedly 
discussed since it contains some basic problems. The Klein-Gordon equation can 
be written for the scalar field $\phi(x)$ with its mass $m$ as
$$ \left({\partial^2\over{\partial t^2}}-\nabla^2 +m^2 \right) \phi (x)= 0  
\eqno{(1.1)}  $$
where we denote $x=(t, \bm{ r})$. The scalar field $\phi$ has only one component, 
and therefore it is believed to describe the spinless particle. The Lagrangian density 
which reproduces eq.(1.1) can be written as
$$  {\cal L} = {1\over 2} \left( \dot{\phi}^\dagger \dot{\phi} -
\nabla{\phi}^\dagger \cdot \nabla{\phi} \right) -
{1\over 2}  m^2{\phi}^\dagger {\phi}  \eqno{ (1.2)} $$
which leads to the Hamiltonian density $\cal H$
$$ {\cal H} = {1\over 2}\left\{  {{\Pi}}^{\dagger}  {\Pi} 
+ \nabla {{\phi}}^\dagger \cdot \nabla {\phi}  +
m^2   {{\phi}}^{\dagger}  {\phi}  \right\}  
\eqno{(1.3)} $$
where ${\Pi} (x)$ is defined as ${\Pi} (x)= \dot{\phi}(x) $. 
In the present days, most of the field 
theory textbooks state that the real scalar field has vanishing current density, but 
if one quantizes the real scalar field, then it can be interpreted as the charge zero 
boson and therefore the zero current density problem may be resolved. 
Further, if the scalar field is complex, then it should correspond to the charged 
bosons like the $\pi^{\pm}$ mesons. This is the standard understanding 
of the real scalar field. 

However, there are basically two serious defects in the real scalar field solution 
in the  Klein-Gordon equation. Zero current density and no non-relativistic limit.  
In this respect, the real scalar field 
is just like a ghost field, but it has been accepted as a physical particle 
and is indeed treated in most of the standard textbooks since it is simple 
in mathematics. 

\section{Klein-Gordon field}
Here, we examine whether a real scalar 
field can exist as a physical observable or not in the Klein-Gordon equation. 
Normally, we find that pion with the positive charge 
is an anti-particle of pion with the negative charge. This can be easily understood 
if we look into the structure of the pion in terms of quarks. $\pi^{\pm}$ are indeed 
anti-particle to each other by changing quarks into anti-quarks. 
Since pion is not an elementary particle, their dynamics must be 
governed by the complicated quark dynamics. Under some drastic approximations, 
the motion of pion may be governed by the Klein-Gordon equation. 

\subsection{Physical Scalar Field}

It looks that eq.(1.1) contains the negative energy state. 
However, one sees that eq.(1.1) is only one component equation and, therefore 
the eigenvalue of $E^2$  can be obtained as a physical observable. 
There is no information from the Klein-Gordon equation for the energy $E$ itself, but 
only $E^2$ as we see it below, 
$$  \left(  -\bm{\nabla}^2+m^2\right) 
\phi = E^2 \phi . \eqno{(2.1)}  $$
In this case, the solution of eq.(2.1) should be described just in the same way 
as the Schr\"odinger field
$$ \phi_{\bm{k}}(x) =A(t) e^{i\bm{k}\cdot \bm{r} } \eqno{(2.2)} $$
which should be an eigenfunction of the momentum operator $\hat{\bm{p}}=-i\bm{\nabla}$.  
The coefficient $A(t)$ can be determined by putting eq.(2.2) into eq.(1.1). 
One can easily find that $A(t)$ should be written as
$$ A(t)={1\over{\sqrt{V\omega_{\bm{k}}}}} a_{\bm{k}} e^{-i\omega_{\bm{k}} t} \eqno{(2.3)} $$
where $a_{\bm{k}}$ is a constant, and 
$\omega_{\bm{k}}$ is given as
$$ \omega_{\bm{k}}=\sqrt{\bm{k}^2+m^2} . \eqno{(2.4)} $$
The shape of $A(t)$ in eq.(2.3) can be also determined from the Lorentz invariance. 
Now, one sees that eq.(2.2) has a right non-relativistic limit. 
This is indeed a physical scalar field solution of the Klein-Gordon equation. 

\subsection{Current Density}
Now, we discuss the current density 
of the Klein-Gordon field which is obtained in terms of the Noether current as 
$$ \rho(x) ={i\over 2}\left(\phi^\dagger(x) {\partial \phi (x)\over{\partial t}}-
{\partial \phi^\dagger(x)\over{\partial t}}\phi(x) \right)  
\eqno{(2.5)} $$
$$ \bm{j}(x) =-{i\over 2}\left[\phi^\dagger(x) (\nabla \phi (x))-
(\nabla \phi^\dagger(x))\phi(x) \right] . 
\eqno{(2.6)} $$

\subsubsection{Classical Real Scalar Field }
The current density of a real scalar field 
vanishes to zero.  This can be easily seen since the real scalar field $\phi (x)$ 
should satisfy 
$$ \phi^\dagger (x)=\phi (x)  .  \eqno{(2.7)} $$
Since the $\phi (x)$ is still a classical 
field, it is easy to see that the current densities of $\rho (x)$ and 
$ \bm{j}(x)$  vanish to zero
$$  \rho(x) = 0, \ \ \ \  \bm{j}(x)=0 . \eqno{(2.8)} $$
This means that there is no flow of the real scalar field, at least, classically. 
This is clear since a real wave function in the Schr\"odinger equation cannot propagate. 
Therefore, the condition that the scalar field should be a real field must be 
physically too strong. In the Schr\"odinger field, one cannot require that 
the field should be real. In fact, if one assumes that the field is real, then one 
obtains that the total energy of  the Schr\"odinger field must vanish, and 
the field becomes unphysical \cite{fujita}. 

Further, if the field is quantized, then 
the current density becomes infinity as shown in ref. \cite{boson}. 

\subsubsection{Physical Scalar Field }
The physical scalar field can be written as in eqs.(2.2) and (2.3)
$$ {\phi} (x) =\sum_{\bm{ k}} {1\over{\sqrt{V\omega_{\bm{ k}}}}}  
a_{\bm{ k}}e^{i{\bm{ k}}\cdot{\bm{ r}}-i\omega_{\bm{ k}} t }. \eqno{(2.9)} $$
In this case, we can calculate the current density 
for the scalar field  with the fixed momentum of ${\bm{ k}}$  
and obtain
$$ \rho(x)= { |a_{\bm{k}}|^2\over V} \eqno{(2.10)} $$
which is positive definite and finite. Therefore, the physical scalar field does not 
have any basic problems. 

\subsection{Composite Bosons}

Pions and $\rho-$mesons are composed of quark and anti-quark fields. Suppressing 
the isospin variables, we can describe the boson fields in terms of the Dirac field 
$ \psi_q (x) \chi_{1\over 2} $ only with the large components, for simplicity
$$ \Psi^{(B)}=\psi_q(x_1) {\psi}_{\bar q}(x_2) 
(\chi^{(1)}_{1\over 2}\otimes \chi^{(2)}_{1\over 2} ) $$
$$=\Phi^{(Rel)}(x) \Phi^{(CM)}(X) \xi_{s,s_z}  \eqno{(2.11)} $$
where $x=x_1-x_2$, $X={1\over 2}(x_1+x_2)$. Here,   
${\psi}_{\bar q}(x) \chi_{1\over 2}$ denotes the anti-particle field and 
$ \xi_{s,s_z}$ is the spin wave function of the boson. 
$\Phi^{(Rel)}(x)$ denotes the internal structure of the boson and 
$\Phi^{(CM)}(X) $ corresponds to the boson field. 
Now, it is clear that the boson field $\Phi^{(CM)}(X) $ is a complex field. 

In the field theory textbooks, the real scalar field 
is interpreted as a boson with zero charge. But this is not the right 
interpretation. The charge is a property of the field in units of the coupling 
constant. The positive and negative charges are connected to the flavor of the scalar 
fields.  For example, a chargeless Schr\"odinger field, of course, has a finite current 
density of $\rho(\bm{r})$. The charge $Q$ of the Schr\"odinger field is given as
$ Q=e_0\int  \rho(\bm{r}) d^3r $ and for the chargeless field, we simply have $e_0=0$, 
which means that it does not interact with the electromagnetic field due to the absence 
of the coupling constant. 

\subsection{Gauge Field}

The electromagnetic field $A_{\mu}$ is a real vector field which is required 
from the Maxwell equation, and therefore it has zero current 
density. However, the gauge field itself is gauge dependent and therefore it is 
not directly a physical observable. 
In this case, the energy flow in terms of the Poynting vector 
becomes a physical quantity. After the gauge fixing and the field quantization, 
the vector field $\hat{\bm{A}}(x)$ can be written as
$$ \hat{\bm{A}}(x)=\sum_{\bm{k}} \sum_{\lambda =1}^2{1\over{\sqrt{2V\omega_{\bm{k}}}}}
\bm{\epsilon}(\bm{k},\lambda) \left[ c_{\bm{k},\lambda} e^{-ikx} +
  c^{\dagger}_{\bm{k},\lambda} e^{ikx} \right]  \eqno{(2.12)} $$
where 
$ \omega_{\bm{k}}=|\bm{k}|$. Here, $ \bm{\epsilon}(\bm{k},\lambda) $ denotes the 
polarization vector. In this case, one-photon state with $(\bm{k},\lambda)$ becomes
$$ \bm{A}_{\bm{k},\lambda}(x)= \langle \bm{k},\lambda|\hat{\bm{A}}(x)|0\rangle 
={1\over{\sqrt{2V\omega_{\bm{k}}}}}
\bm{\epsilon}(\bm{k},\lambda) e^{-i{\bm{ k}}\cdot{\bm{ r}}+i\omega_{\bm{ k}} t }  $$
which is the eigenstate of the momentum operator $\hat{\bm{p}}=-i\bm{\nabla}$. 
In this respect, the gauge field 
$A_{\mu}$ is completely different from the Klein-Gordon scalar field. Naturally, 
the gauge field  does not have any corresponding non-relativistic field.

\section{Boson in the Schwinger model}

There is a good example of the physical boson which is described in terms 
of the fermion operators in the Schwinger model. Since the boson in the Schwinger 
model is described in terms of the physical quantities, we can learn the essence 
of the physical boson how nature makes up a boson. Also, we learn the field 
quantization procedure of the boson fields.

\subsection{Schwinger model}
Here, we first describe briefly the Schwinger model which is a two dimensional QED 
with massless fermion. Its Lagrangian density can be written as \cite{schw}
$$  {\cal L} =  \bar \psi i \gamma_{\mu} D^{\mu}  \psi
  -{1\over 4} F_{\mu\nu} F^{\mu\nu}   \eqno{ (3.1)} $$
where
$$  D_{\mu}= {\partial}_{\mu} + ig A_{\mu}, \quad
  F_{\mu\nu}= \partial_{\mu} A_{\nu} - \partial_{\nu} A_{\mu} . 
 \eqno{ (3.2)}    $$
Here, $A_{\mu}$ denotes the vector potential in two dimensions. 
By taking the Coulomb gauge fixing $\partial^1 A_1 =0 $, we can describe 
the Hamiltonian of massless QED$_2$  
$$ H=\int  \bar{\psi}(x)  \left[-i {\gamma^1}\partial_1
+g {\gamma^1}{A_1} \right]\psi(x)  dx +{1\over 2}\int \dot{A}_1^2 dx $$
$$ -{g^2\over 4} \int j_0(x) |x-x'| j_0(x')dxdx'  
\eqno{ (3.3)}  $$
where the fermion current $j_\mu(x) $ is defined as
$$ j_\mu(x) = \bar{\psi}(x) \gamma_\mu \psi(x) . $$
The massless QED in two dimensions has a mass scale, that is,
the coupling constant $g$ has a mass dimension. Therefore, all the physical
observables such as a boson mass are measured by the coupling constant $g$.

\subsection{Bosonization}
The Hamiltonian of the Schwinger model can be bosonized in the following fashion 
\cite{manton}
$$ H= {1\over 2}\sum_{p} \left\{  {\tilde{\Pi}}^{\dagger} (p) \tilde{\Pi} (p)
+ p^2 {\tilde{\Phi}}^\dagger (p)\tilde{\Phi} (p) +
{\cal M}^2   {\tilde{\Phi}}^{\dagger} (p) \tilde{\Phi} (p) \right\} 
\eqno{(3.4)} $$
where the mass of the boson (Schwinger
boson ${\cal M}$) is given as
$$ {\cal M}={g\over{\sqrt{\pi}}} . $$
The boson fields $\tilde{\Phi}(p)$ and its conjugate field $\tilde{\Pi}(p)$ 
are related to the fermion 
current density in momentum representation $ \tilde{j_{0} }(p) $ and 
$ \tilde{j_{1} }(p) $ \cite{fujita}
$$ \tilde{j_{0} }(p) =ip\sqrt{4\pi\over L} \tilde{\Phi} (p)=\int j_0(x) e^{ipx} dx 
 \eqno{(3.5a)} $$
$$ \tilde{j_{1} }(p) =\sqrt{L\over \pi} \tilde{\Pi} (p) =\int j_1(x) e^{ipx} dx 
\eqno{(3.5b)} $$
where $\tilde{\Phi} (p)$ and $\tilde{\Pi} (p)$ satisfy the bosonic commutation relation
$$ [\tilde{\Phi} (p), \tilde{\Pi}^\dagger (p')]= i\delta_{p,p'} . \eqno{(3.6)} $$
This commutation relation is proved under the condition that the fermion operators 
$ \tilde{j_{0} }(p) $ and $ \tilde{j_{1} }(p)$ should operate on the physical states 
where, in the deep negative states, all the states are occupied while, in the highly 
excited states, there is no particle present.   

For the zero mode, fields $\tilde{\Phi}(0)$ and its conjugate field $\tilde{\Pi}(0)$ are 
related to the regularized chiral charge and its time derivative, but we do not write 
them here since they are not relevant to the present discussion.  

\subsection{Complex field}
The boson field $\tilde{\Phi}(p)$ is a complex function as one sees it from eq.(3.5) and 
can be written in terms of $\phi(x)$ in eq.(2.9)
$$ \tilde{\Phi}(p) ={1\over{\sqrt{L}}} \int \phi(x)  
e^{-ipx+i\omega_p t}dx ={1\over{\sqrt{\omega_p}}}c_p . \eqno{(3.7)}  $$
Using $ \Pi(x) = \dot{\phi}(x) $, we find 
$$ \tilde{\Pi} (p)=-i{\sqrt{\omega_p}}c_p . \eqno{(3.8)}  $$
In this case, one sees that the operators $c_p$, $c^\dagger_p$ 
satisfy the right commutation relation due to eq.(3.6)
$$ [c_p, c^\dagger_q ] = \delta_{p,q} . \eqno{(3.9)} $$
Therefore, we can write the Hamiltonian of eq.(3.4) as
$$ H=\sum_p \omega_p c^\dagger_p c_p \eqno{(3.10)}  $$
which is just the proper expression of the boson Hamiltonian. It should be 
interesting to note that the boson Hamiltonian of eq.(3.10) has no zero point 
energy which is normally an infinite quantity. This must be related to the fact 
that the original Hamiltonian which is written in terms of the fermion operators 
does not contain any zero point energy. In fact, in the bosonization procedure, 
the vacuum energy as well as the charges of fermions are regularized in the 
gauge invariant fashion. 

From this example of the physical boson field, we see that the scalar 
Klein-Gordon field must be described by the complex field which corresponds to one flavor 
boson. If one wishes to describe two charged bosons, then one has to introduce the 
isospin corresponding to a new degree of freedom. It is by now clear that 
there should not exist any real scalar field as a physical observable. 

\section{Higgs Mechanism}

If the scalar field is a complex function, then many properties 
of bosons are physically acceptable. In this case, however, there is a serious problem 
in connection with the Higgs mechanism. 
The Lagrangian density of the complex scalar field $\phi(x)$  which
interacts with the $U(1)$ gauge field can be written as
$$  {\cal L} = {1\over 2}  (D_\mu\phi)^\dagger (D^\mu\phi) -u_0 \left(  |{\phi}|^2
-\lambda^2 \right)^2 -{1\over 4} F_{\mu\nu} F^{\mu\nu}   \eqno{ (4.1)} $$
where $u_0$ and $\lambda$ are constant. 
In the Higgs mechanism, the scalar field is rewritten in terms of the two real 
scalar fields $\eta(x)$ and $\xi(x)$  as
$$ \phi(x) = \left(\lambda + \eta(x)\right) e^{i{\xi(x)\over{\lambda}}}.  \eqno{(4.2)} $$
After the spontaneous symmetry breaking, the $\eta(x)$ field remains physical, and 
the $\xi(x)$ field is absorbed into the gauge field which becomes massive by acquiring 
one degree of freedom. 
However, since the real scalar field is not physical, it is quite difficult 
to carry out the symmetry breaking mechanism in this fashion. 
At least, one cannot understand what the real scalar field indicates after 
the Higgs mechanism since the Lagrangian density  
contains the real scalar field of $\eta(x)$ after the spontaneous symmetry breaking. 

In this respect, one should reexamine the Higgs mechanism from the point of view of 
the non-existence of the real scalar field. There is no doubt that the Weinberg-Salam 
model has achieved a great success for describing many experimental observations 
in connection with the weak decays. Nevertheless, there are still some fundamental 
questions left for the basic mechanism of the spontaneous symmetry breaking 
on which the Weinberg-Salam model is entirely dependent. 


\section{Discussions}

The Klein-Gordon equation was discovered 80 years ago, and since then 
the boson is believed to be described by the Klein-Gordon equation. 
The scalar boson should exist if it is a composite object. 
In this case, one sees that the boson field should be complex like the Schr\"odinger 
field. In most of the field theory textbooks, however, it has been well accepted that 
the real scalar field should physically exist and that the real scalar field is 
described as the chargeless particle. At the same time, people realized that the real 
scalar field has some problems like vanishing current density. Therefore, it is stated 
in the textbooks that the real scalar field should be always quantized, and then 
it is all right. 

Here, we point out that there should not exist any real scalar field. Instead, 
the scalar field must be always a complex function with only one component. 
In this case, one can define the finite current density and also the scalar 
field can possess the proper non-relativistic limit when the motion is slow. 
In addition, the bosonization procedure of the Schwinger model shows that the boson 
is a complex field with one flavor. Since the Schwinger  boson is a physical object, 
it strongly suggests that the scalar field should be a complex field. 

However, it is still not yet settled whether the scalar field or the Klein-Gordon 
equation itself for the elementary fields is physically acceptable or not. 
If the Klein-Gordon equation is fundamental, then it should have a proper degree of 
freedom which should be two while the Dirac equation has a correct degree of freedoms 
as a fermion field which is four. 

The derivation of the Klein-Gordon equation is closely connected with the first 
quantization procedure which should be understood more in depth \cite{fujita,fks2}.


\begin{thebibliography}{99}




\bibitem{fujita}
T. Fujita, "Symmetry and Its Breaking in Quantum Field Theory", 
(Nova Science Publishers, 2006) 

\bibitem{boson}
S. Oshima, S. Kanemaki and T. Fujita, "Problem of Real Scalar Klein-Gordon Field", 
hep-th/0512156.

\bibitem{schw}
J. Schwinger, \emph{Phys. Rev.} \textbf{128}, 2425 (1962)


\bibitem{manton}
N. S. Manton, \emph{Ann. Phys.} \textbf{159}, 220 (1985)


\bibitem{fks2}
T. Fujita, S. Kanemaki and S. Oshima, \emph{``New Concept of First
Quantization"}, hep-th/0601102.


\end{thebibliography}
\end{document}